\documentclass[floatfix,aps,twocolumn,superscriptaddress,prl]{revtex4-1}
\usepackage{graphicx}
\usepackage{dcolumn}
\usepackage{bm}
\usepackage[T1]{fontenc}
\usepackage[latin9]{inputenc}
\usepackage{textcomp}
\usepackage{amsmath}
\usepackage{mathrsfs}
\usepackage{graphicx}
\usepackage{amssymb}
\usepackage{units}
\usepackage{xcolor}
\usepackage{float}

\begin{document}

\title{Two-dimensional polaron spectroscopy of Fermi superfluids}

\author{Ivan Amelio}
\affiliation{
Institute of Quantum Electronics ETH Zurich, CH-8093 Zurich, Switzerland}

\begin{abstract}
Multidimensional spectroscopy is becoming an increasingly popular tool and  there is an ongoing effort to access electronic transitions and many-body dynamics in correlated materials.
We apply the protocol recently proposed by Wang~\cite{wang2022multidimensional}  to extract
two-dimensional polaron spectra  in a Fermi superfluid with an impurity. The bath is descibed by a BCS ansatz and it assumed that the impurity can scatter at most one quasiparticle pair.
The spectral response contains a symmetric contribution, which carries the same information as Ramsey spectra, and an asymmetric one.
While {\it a priori} it may seem promising to probe the quasiparticle gap from the asymmetric contribution, we show explicitly that this is not the case
and, in the absence of incoherent processes, multidimensional spectroscopy does not bring much additional information.
Our calculation is suitable for 3D ultracold gases, 
but we discuss implications for exciton-polarons in 2D materials.
\end{abstract}

\date{\today}
\maketitle

{\em Introduction.}
Multidimensional spectroscopy~\cite{jonas2003two,li20171chapter,Gelzinis2019two-dimensional} is an experimental tool that allows to study  extremely fast processes with high spectral resolution. It has been tremendously successful in investigating the mechanisms underlying photosynthesis~\cite{schlaucohen2011two-dimensional}  and it is an essential tool to understand the incoherent and coherent energy transfer in molecular aggregates and the dynamics electronic transitions.
Currently, there is growing interest in using this approach to explore the many-body properties of correlated materials~\cite{Beaulieu2021,valmilspild2022strongly},
including cuprates.
Onging attempts to extend the technique to the terahertz range are also worth to be mentioned~\cite{Lu2018two-dimensional}.

Recently,
Wang proposed \cite{wang2022multidimensional,wang2022twodimensional} an extension of multidimensional spectroscopy suitable for cold atoms experiments, consisting in immersing an impurity in a Fermi sea and performing a  sequence of four Rabi pulses.
Here, we will refer to this approach as to two-dimensional polaron spectroscopy (2DPS), since the dressing of the impurity by the excitations  of the bath to form polaronic states plays a central role for the impurity dynamics.

In this Letter we are interested in applying the 2DPS protocol to an impurity immersed in a three-dimensional Fermi superfluid, where spin up fermions pair with spin down fermions~\cite{ferrier2014mixture}.  Polaron formation along the BEC-BCS crossover is in itself a topic of intense theoretical research~\cite{nishida2014polaronic,yi2015polarons,hu2022crossover,wang2022exact,wang2022heavy,?}. The main challenge is to provide an accurate description both in the  BCS and  BEC regime, to recover the Fermi and Bose polaron cases respectively. 

Using a generalized Chevy ansatz~\cite{chevy2006universal} on top of a BCS variational state~\cite{engelbrecht1995}, we compute the Ramsey one-dimensional spectra as well as the 2DPS ones.
We show that, in the absence of incoherent energy transfer, the 2DPS does not bring much additional information. In particular, there is no direct signature of the quasiparticle gap in the asymmetric contribution to the 2DPS.
Our results may help to elucidate experimental data in future studies of multidimensional spectroscopy in superconductors. 

~

{\em Polarons in  Fermi superfluids.}
We consider a zero temperature gas of spin one-half fermionic atoms  described by the annihilation operators $c_{\mathbf{k}\sigma}$, where $\sigma=\uparrow,\downarrow$, and by the dispersion 
$\xi_k = \frac{k^2}{2m} - \mu$,
where  $m$ is the mass and $\mu$ the chemical potential.
A single impurity of mass $M$ is also present and it has two internal states, 
splitted by a large energy $\omega_0$ and described by $d_{\mathbf{k}\sigma}$.
Each fermion interacts with a spin-flipped atom via a contact attractive interaction with coupling strength $g$. This is related to the scattering length $a$ in the usual way
$\frac{1}{g} = \frac{m}{4\pi a} - \frac{1}{V} \sum_{\mathbf{k}}^\Lambda \frac{m}{k^2}$,
where $V$ is the volume of the system and $\Lambda$ a cutoff.
We assume that the impurity interacts only with the  $\uparrow$ fermions and only when in its $\uparrow$ state,
with strength $g_\uparrow$ and scattering length $a_\uparrow$. This is reasonable experimentally,  since  Feschbach resonances  involve a specific spin configuration.

The overall Hamiltonian reads
\begin{multline}
H = 
   \sum_{\mathbf{k}\sigma} 
   \xi_k
   c^\dagger_{\mathbf{k}\sigma} c_{\mathbf{k}\sigma} + \frac{g}{V} \sum_{\mathbf{k}\mathbf{p}\mathbf{q}} c^\dagger_{\mathbf{k}+\mathbf{q}\uparrow}  c^\dagger_{\mathbf{p}-\mathbf{q}\downarrow} c_{\mathbf{p}\downarrow} c_{\mathbf{k}\uparrow}
   + 
   \\
   +
    \sum_{\mathbf{k}\sigma} 
   \left(
\omega_{0}\delta_{\sigma\uparrow} + \frac{k^2}{2M} 
   \right)
   d^\dagger_{\mathbf{k}\sigma} d_{\mathbf{k}\sigma} + \frac{g_\uparrow}{V} \sum_{\mathbf{k}\mathbf{p}\mathbf{q}} c^\dagger_{\mathbf{k}+\mathbf{q}\uparrow}  d^\dagger_{\mathbf{p}-\mathbf{q}\uparrow} d_{\mathbf{p}\uparrow} c_{\mathbf{k}\uparrow}
\end{multline}
and it is actually convenient to split it as 
$H=H_\uparrow + H_\downarrow$
depending on the internal state of the impurity.

Pairing in the ground-state of the bath can be 
qualitatively captured by the variational  BCS ansatz
along all the BEC-BCS crossover \cite{engelbrecht1995}. A convenient approach is to introduce the fermionic quasiparticle operators 
\begin{equation}
    \gamma_{\mathbf{k}\uparrow} = u_k c_{\mathbf{k}\uparrow} + v_k c_{-\mathbf{k}\downarrow}^\dagger
    \ , \ \ \ \ \ \
        \gamma_{\mathbf{k}\downarrow} = u_k c_{\mathbf{k}\downarrow} - v_k c_{-\mathbf{k}\uparrow}^\dagger \ .
\end{equation}
We define $|\text{BCS}\rangle$ as the vacuum of the quasi-particles  
$
    \gamma_{\mathbf{k}\sigma} |\text{BCS}\rangle = 0.
$
 Writing  $u_k = \cos\theta_k, v_k = \sin\theta_k$, the variational parameter $\theta_k$ minimizes the energy for
$
    \tan2\theta_k = -
    \frac{\Delta}{\xi_k},
$
where 
$ 
\Delta = \frac{g_\uparrow}{V}  \sum_\mathbf{k} 
\langle  c_{\mathbf{k}\uparrow} c_{\mathbf{k}\downarrow} \rangle_{\text{BCS}}
=
\frac{-g_\uparrow}{V}  \sum_\mathbf{k} 
  u_k v_k    
$ 
is the pairing order parameter.
In the following we use 
$E_F = \frac{k_F}{2m}
=
\frac{3\pi^2}{m} n
$
as unit, which fixes $k_F$ and the density of each spin $n = \frac{1}{V}\sum_\mathbf{k}   \langle  c_{\mathbf{k}\sigma}^\dagger c_{\mathbf{k}\sigma} \rangle_{\text{BCS}}$, where $\langle  c_{\mathbf{k}\sigma}^\dagger c_{\mathbf{k}\sigma} \rangle_{\text{BCS}} = v_k^2$.

Exciting a  quasiparticle costs on average an energy
$E_k = \sqrt{\xi_k^2 + \Delta}$, meaning that
$\langle \gamma_{\mathbf{k}\sigma} H \gamma_{\mathbf{k}\sigma}^\dagger \rangle_{\text{BCS}}  =
    E_k$, having shifted the zero of the energy so to have $\langle  H \rangle_{\text{BCS}}=0$.
One crucial point is that we do not approximate the Hamiltonian of the bath to 
$H_{\text{BCS}}
=
\sum_{\mathbf{k}\sigma} E_k
\gamma_{\mathbf{k}\sigma}^\dagger \gamma_{\mathbf{k}\sigma}
$,
but we retain the interactions between the quasiparticles. As illustrated in details in [???], this  entails that, rather than a large gap $2\Delta$, the excitation modes in the 2-quasiparticle Hilbert sector have a small gap, stemming from (an imprecise description of) the gapless Goldstone mode.

With this in mind, 
the impurity problem is diagonalized in the 
2-quasiparticle Hilbert subspace.
The idea is that the impurity can scatter the bosonic excitations of the system, and in this case we restrict to a single excitation to make the problem treatable.
In other words, the polaron wavefunctions are given by the generalized Chevy ansatz~\cite{yi2015polarons,hu2022crossover,?}
\begin{equation}
    |\Psi\rangle 
    =
    \left(
\psi_0 d^\dagger_{\mathbf{0}\uparrow} + \frac{1}{V}
\sum_{\mathbf{k}\mathbf{Q}} 
\psi_\mathbf{k}(\mathbf{Q})
d^\dagger_{\mathbf{Q}\uparrow}
\gamma^\dagger_{\mathbf{k}\downarrow}
\gamma^\dagger_{\mathbf{-k-Q}\uparrow} 
    \right)
    |\text{BCS}\rangle,
\end{equation}
where here we restrict to total zero momentum and to the sector with the impurity in the interacting internal state.
Notice that this ansatz contains exactly the 3-body bound state in the vacuum.
The Schr\"odinger equations in the variational subspace read
\begin{equation}
    i\partial_t \psi_0 = 
    \frac{g_\uparrow}{V^2} \sum_{\mathbf{k}\mathbf{Q}}
    v_\mathbf{k} u_{\mathbf{k}+\mathbf{Q}} \psi_\mathbf{k}(\mathbf{Q})
    \label{eq:SChevy0}
\end{equation}
\begin{multline}
   i\partial_t  \psi_\mathbf{k}(\mathbf{Q})
   =
   \left(
   E_\mathbf{k} + E_{\mathbf{k}+\mathbf{Q}} + \omega_0 
+ \frac{Q^2}{2M} 
\right) \psi_\mathbf{k}(\mathbf{Q})
      +
   \\
   + g_\uparrow v_\mathbf{k} u_{\mathbf{k}+\mathbf{Q}} \psi_0 
   +
   \frac{g_\uparrow}{V} \sum_{\mathbf{Q'}}
    u_{\mathbf{k}+\mathbf{Q}} u_{\mathbf{k}+\mathbf{Q'}} \psi_\mathbf{k}(\mathbf{Q'})
    +
    \\
    +
   \frac{g}{V}
   u_{\mathbf{k}} u_{\mathbf{k}+\mathbf{Q}}
   \sum_{\mathbf{k'}}
    u_{\mathbf{k'}} u_{\mathbf{k'}+\mathbf{Q}} \psi_\mathbf{k'}(\mathbf{Q})
    \label{eq:SChevykQ}
\end{multline}
In writing Eqs.~(\ref{eq:SChevy0}-\ref{eq:SChevykQ}) we used the fact that for contact interactions terms like $\frac{g}{V}\sum_\mathbf{k} ... v_k$ are subleading as $V \to \infty$.
For the impurity in the non-interacting $\downarrow$ state, one can  factorize the impurity wavefunction, while quasiparticles interact with each other.
Below we will use the matrix form
$
    H_\sigma = V_\sigma E^\sigma V_\sigma^\dagger
$
for the eigenmodes and eigenergies in the
Chevy subspace, having diagonalized the Hamiltonian in the non-interacting and interacting impurity sectors. Here $E^\sigma$ is a diagonal matrix 
and we index the Chevy subspace basis in such a way that the state $j=0$ is the state with no quasiparticles $\psi_0 d^\dagger_{\mathbf{0}\sigma}
    |\text{BCS}\rangle$. Also, in the $\sigma=\downarrow$ sector this is the 0-th eigenstate with energy $E^\downarrow_0=0$.

\begin{figure}[t]
    \centering
    \includegraphics[width=0.48\textwidth]{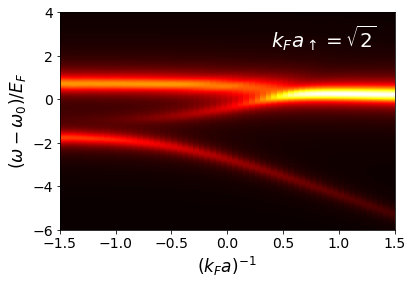}
    \caption{The one-dimensional Ramsey spectrum is plotted as a function of the fermion-fermion scattering length $a$, for fixed fermion-impurity interactions $a_\uparrow$. The left (right) regions correspond to the BCS (BEC) regime.
    }
    \label{fig:1d_spectrum}
\end{figure}

In experiments a typical protocol is  Ramsey spectroscopy~\cite{goold2011orthogonality,knap2012time-dependent,adam2022coherent}, which consists in starting with the non-interacting impurity and applying two $\pi/2$ pulses separated by a time $t$.  Varying $t$  allows to probe the response function in frequency space
\begin{equation}
    \mathcal{A}(\omega) = 
    \sum_m \frac{|(V_\uparrow)_{0m}|^2}{\omega - E_m^{\uparrow} + i0^+}
\end{equation}

The one-dimensional spectrum 
$-2\text{Im}\mathcal{A}(\omega)$
of the system  for the interacting impurity is plotted in Fig.~\ref{fig:1d_spectrum} 
for different fermion-fermion scattering length $a$
for $k_F a^\uparrow = \sqrt{2}$, which corresponds to a fermion-impurity binding energy of $E_F$ in the vacuum. A broadening $0^+ \to 0.3 E_F$ is used in practice.

The two main features are a lower and upper branch, denoted attractive (AP) and repulsive (RP) polarons respectively.
The AP state is closely linked to the bound state of the impurity with a fermion or a fermion pair, while the RP is continuosly connected to the bare impurity.
As already observed in \cite{yi2015polarons} a secondary middle peak is present on the BCS side of the crossover; consistently with the symmetry of the wavefunction and the size of the gap, we attribute this feature to a Cooper pair excited into the Higgs channel bound to the impurity.
 The divergence of the AP energy in the deep BEC limit is explained by  
 the dependence of the effective dimer-impurity scattering length with $a$
 in the three-body problem \cite{cui2014atom-dimer,zhang2014calibration}.
 We also recall that the cutoff $\Lambda$ cannot be removed in the three-body problem with contact interactions (Thomas collapse). Here we use $\Lambda = 20 k_F$.

\begin{figure*}[t]
    \centering
    \includegraphics[width=0.98\textwidth]{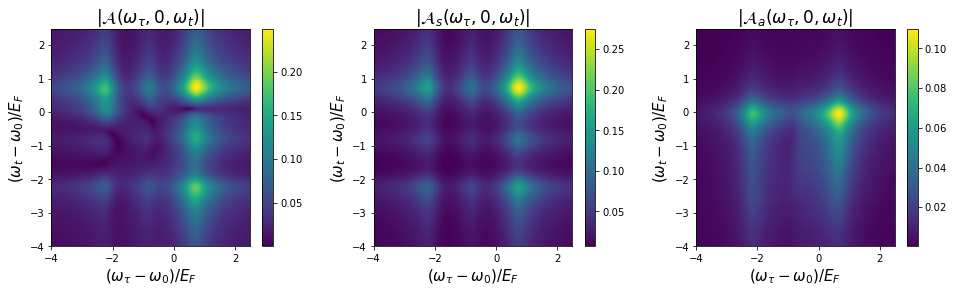}
    \caption{
    The modulus of the 2DPS response function is shown for $(k_F a)^{-1} = -0.5, (k_F a)^{-1} = 1/\sqrt{2}$ and $T=0$.
    More precisely, from left to right, we display the total spectrum, the symmetric contribution and the asymmetric one. The lower (upper) peak on the diagonal corresponds to the AP (RP).
    }
    \label{fig:trittico}
\end{figure*}

~

{\em Two-dimensional polaron spectroscopy (2DPS).}
Taking inspiration on multidimensional spectroscopy~\cite{Gelzinis2019two-dimensional},
Wang has recently proposed~\cite{wang2022multidimensional} a generalization of the Ramsey protocol. This new 2DPS approach consists in applying four $\pi/2$ pulses to  the two-level impurity atom and measuring its final state. The three time intervals between the four pulses are denoted in order coherence time $\tau$, waiting time $T$ and detection time $t$.
The two-dimensional spectrum is obtained Fourier transforming with respect to $\tau$ and $t$.

Assuming that $\omega_0$ is a very large energy scale compared to the AP and RP resonances, one can invoke a rotating-wave approximation and write the response function
$\mathcal{A}(\tau,T,t)  = \frac{I_1 + I_2}{2}$, where the two contributions are 
\begin{equation}
I_1 =
    \langle i |
e^{iH_\downarrow\tau} s_-
e^{iH_\uparrow(T+t)} s_+
e^{-iH_\downarrow t} s_-
e^{-iH_\uparrow(T+\tau)} s_+
    | i \rangle
    \label{eq:I1}
\end{equation}
\begin{equation}
    I_2 =
    \langle i |
e^{iH_\downarrow(\tau+T)} s_-
e^{iH_\uparrow t} s_+
e^{-iH_\downarrow(t+T)} s_-
e^{-iH_\uparrow\tau} s_+
    | i \rangle
    \label{eq:I2}
\end{equation}
with  $s_+ = s_-^\dagger = \sum_{\mathbf{k}} d^\dagger_{\mathbf{k}\uparrow} d_{\mathbf{k}\downarrow}$
the Rabi flip operator
and
$| i \rangle = d^\dagger_{\mathbf{k}\downarrow} |\text{BCS} \rangle$ the initial state.

The spectral function can be easily evaluated diagonalizing the Hamiltonian in the Chevy subspace.
After a double Fourier transform and using  $E^\downarrow_0=0$, one gets
\begin{multline}
    \mathcal{A}(\omega_\tau,T,\omega_t) 
    =
    \frac{1}{2\pi^2} \sum_{j mn}
    (e^{-(E^\uparrow_n-E^\uparrow_m)T} + 
    e^{-iE^\downarrow_j  T}) 
    \times
    \\
    \times
    \frac{(V_{\uparrow})_{0m}(V_\uparrow^\dagger V_\downarrow)_{mj}}{
\omega_t - E^\uparrow_m + E^\downarrow_j
- i0^+
    } \ 
    \frac{(V^*_{\uparrow})_{0n}(V_\downarrow^\dagger V_\uparrow)_{j n}}{
\omega_\tau - E^\uparrow_n + i0^+
    }.
\end{multline}
This expression generalizes the result of \cite{wang2022twodimensional} for the Fermi polaron, where it holds $(V_{\downarrow})_{jm} = \delta_{jm}$.

The symmetric contribution $\mathcal{A}_s$ is defined by restricting the summation over $j=0$ and, since $E^\downarrow_0=0$, it obeys
$\mathcal{A}_s(\omega_\tau,T,\omega_t) = \mathcal{A}_s^*(\omega_t,T,\omega_\tau)$. It is clear that this term does not contain additional information with respect to the Ramsey spectrum $\mathcal{A}(\omega)$, in particular $\mathcal{A}_s(\omega_\tau,0,\omega_t)
\propto \mathcal{A}(\omega_\tau)\mathcal{A}(\omega_t)$
for $T=0$
and the quantum oscillations of the cross-peaks with $T$ merely reflect the energy difference between the resonances in $\mathcal{A}(\omega)$.

\begin{figure}[t]
    \centering
    \includegraphics[width=0.48\textwidth]{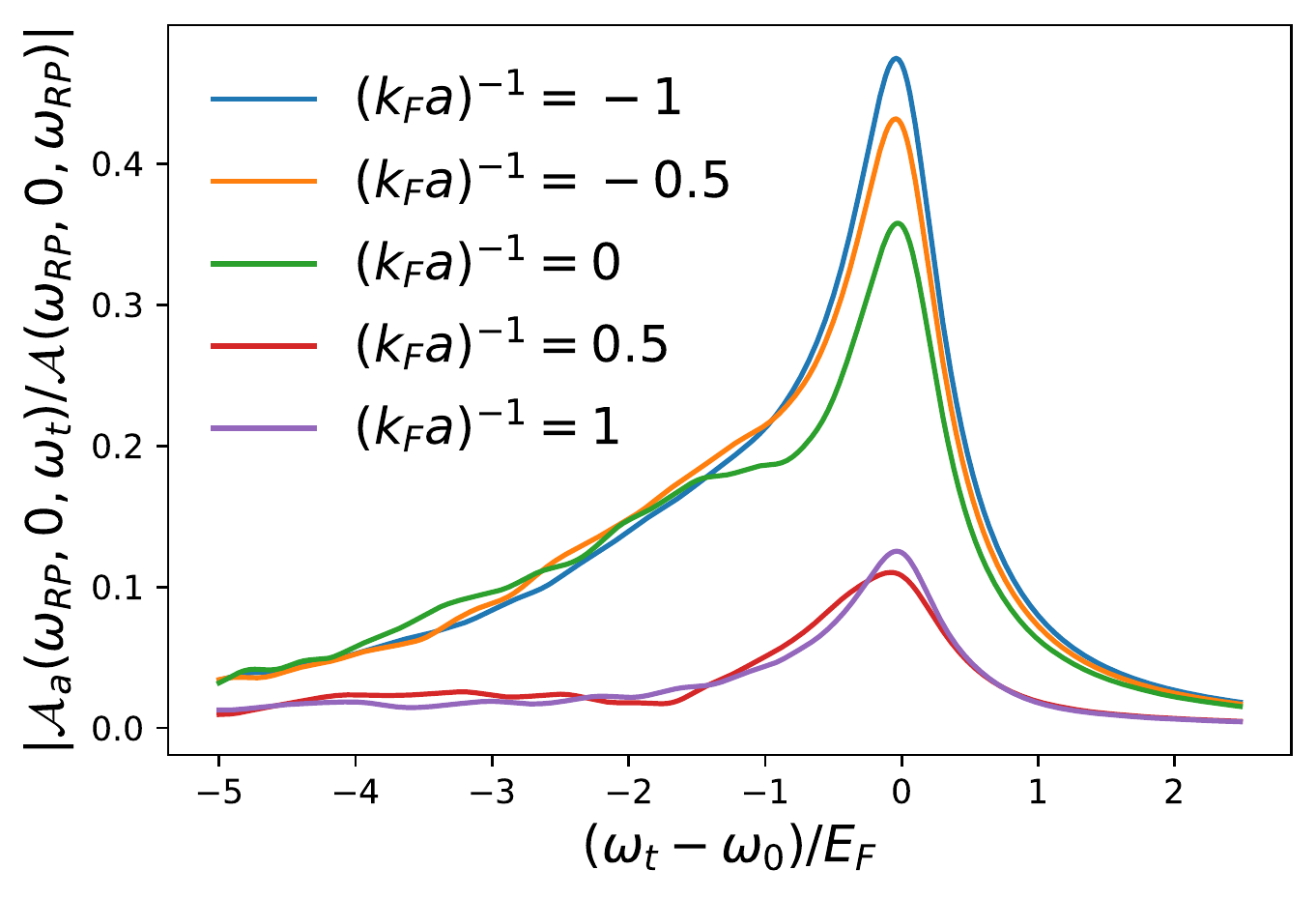}
    \caption{The asymmetric part of the  2DPS is plotted along the  $T=0, \omega_\tau = \omega_{RP}$ slice and at different $k_F a$. More precisely, the intensity is normalized to the RP diagonal peak. The asymmetric contribution is always peaked around $\omega_0$ and decreases approaching the BEC limit of the crossover. 
    }
    \label{fig:asymmetric}
\end{figure}

\begin{figure}[t]
    \centering
    \includegraphics[width=0.45\textwidth]{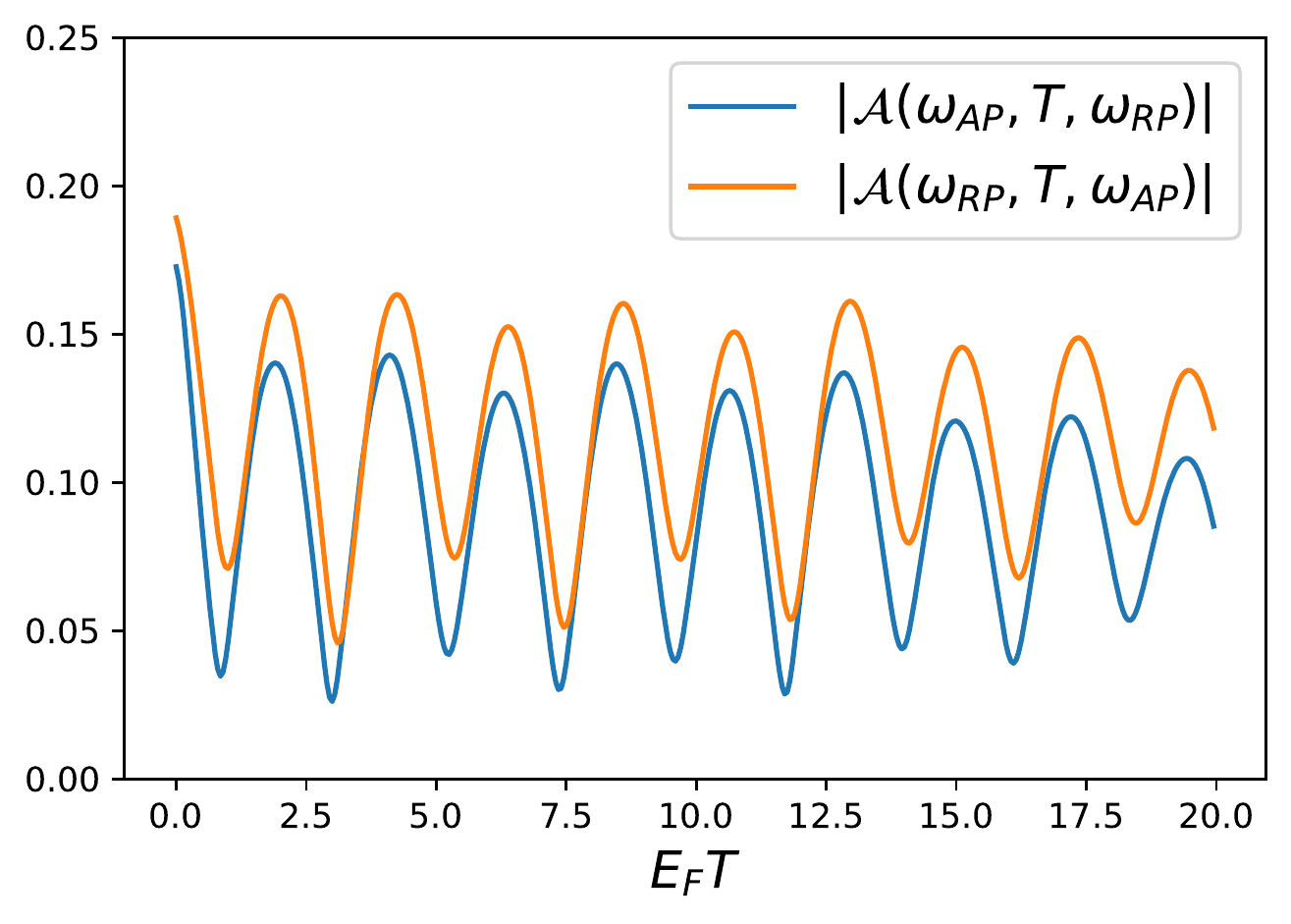}
    \caption{
    The oscillations of the two main cross-peaks are plotted as a function of the waiting time $T$.
    }
    \label{fig:quantum_oscillations}
\end{figure}

Looking at the expression for the asymmetric term, instead, one can hope to extract some intrinsic  properties of the bath in the absence of the impurity. Indeed, setting for simplicity $T=0$, one notices that the $m$-th pole is shifted to  $E^\uparrow_m - E^\downarrow_j$.
In the Fermi polaron problem analysed in \cite{wang2022twodimensional}, the asymmetric term leads to a characteristic shoulder in the two-dimensional spectrum, at a detection frequency below the RP resonance.

While the particle-hole excitation spectrum of a Fermi sea is gapless, for a Fermi superfluid the 2-quasiparticle continuum has a gap $2\Delta$, as a consequence of the fact that introducing a fermion in the system costs a finite energy. We know that actually the bosonic excitation spectrum is  gapless, since it doesn't cost any energy to add a Cooper pair to the condensate, but it is plausible that the 2-quasiparticle continuum has a major effect in the 2DPS, because of its large density of states. 
Therefore, 2DPS seems {\em a priori}
a promising tool to probe the gap of a Fermi superfluid. In particular, one would expect the asymmetric term to yield some feature at a detection frequency $\sim 2\Delta$ below the RP resonance.
We show below that unfortunately this is not the case.

The 2DPS spectra are reported in Fig.~\ref{fig:trittico} for $(k_F a)^{-1} = -0.5, (k_F a)^{-1} = 1/\sqrt{2}$ and zero waiting time.  
In the left panel we plot the modulus of the total response function  $|\mathcal{A}(\omega_\tau,0,\omega_t)|$, and in the other two panels the symmetric and asymmetric contributions are shown.
Spectra for other points along the BEC-BCS crossover as well as the real part of $\mathcal{A}$ can be found in the SI. 
The main features of Fig.~\ref{fig:trittico} are nine peaks arising from the three resonances already present in the Ramsey spectra of Fig.~\ref{fig:1d_spectrum}.

The other visible feature is the destructive interference dip visible around $(\omega_\tau, \omega_t) \sim (\omega_{RP},  \omega_0)$ and which can clearly be ascribed to the asymmetric term.
In the absence of incoherent processes, the main difference between Ramsey spectroscopy and 2DPS is this dip. We then study the asymmetric term, fixing $\omega_\tau=\omega_{RP}$ and $T=0$ for different fermion-fermion scattering length. 
In Fig.~\ref{fig:asymmetric} we normalize the asymmetric contribution to the height of the RP diagonal peak. For every value of the scattering length, the ratio $|\mathcal{A}_a(\omega_{RP}, 0, \omega_t)/\mathcal{A}(\omega_{RP}, 0, \omega_{RP})|$ peaks at $\omega_t \sim \omega_0$ and becomes weaker on the BEC side of the crossover.

We also plot the so-called quantum oscillations of the AP-RP and RP-AP cross-peaks. The carrier frequency corresponds to $\omega_{RP}-\omega_{AP}$, while the gentle envelope behavior is due to the middle resonance. 
Naively, one could have instead expected $\omega_{RP}-\omega_{AP}+2\Delta$ for the carrier frequency.

The absence of a clear signature of the size of the pairing gap is, as mentioned before, due to the fact that one has to consider interacting quasiparticles which form gapless Cooper pairs. If one approximates $H_\downarrow \simeq H_{BCS} +  \sum_{\mathbf{k}\sigma} 
  \frac{k^2}{2M} 
   d^\dagger_{\mathbf{k}\sigma} d_{\mathbf{k}\sigma}$, one would indeed get spurious  peaks in $\mathcal{A}_a$ not at $\omega_0$, as demonstrated in the SI.

~

{\em Discussion and conclusions.}
In this Letter we have computed the 2DPS spectra for an impurity immersed in a three-dimensional Fermi superfluid. Our results suggest that this tool would not be very useful in measuring directly the size of the gap and, in the absence of incoherent energy transfer,  little information is added with respect to the standard Ramsey spectroscopy. 

While to our knowledge in cold atom experiments   Wang's protocol has not been implemented yet, two-dimensional spectroscopy of solid-state samples is by now an established tool~\cite{hao2016coherent,helmrich2021phonon-assisted, wang2022excited}, which allows to access the third order nonlinear susceptibility $\chi(\tau,T,t)$.
The formalism reviewed here is particularly suitable to be applied to transition metal dichalcogenide heterostructures of few layers~\cite{mak2016photonics}.
The optical response of these semiconductors is dominated by excitons. In the presence of some doping, the excitons are dressed by the electronic particle-hole excitations and form exciton-polarons~\cite{mak2013tightly,sidler2017fermi}.
In the expressions (\ref{eq:I1},\ref{eq:I2}) for $I_1,I_2$ one should replace 
$s_+$ with the exciton creation operator $x^\dagger_\mathbf{0}$, where the subscript indicates that the exciton momentum is very small, since it should belong to the light-cone. 
The two inner excitonic operators then add an extra projection of the impurity in the zero momentum state.
As a consequence, the asymmetric contribution $\mathcal{A}_a$ 
is suppressed by a factor of the order $\frac{a^2_T}{\lambda^2} \sim 10^{-4}$, where the trion radius $a_T$ determines the typical scale of the momentum states  involved in the AP and RP peaks.
This argument is perfectly consistent with the experimental findings of \cite{hao2016coherent}, that have been fitted neglecting the asymmetric contribution~\cite{tempelaar2019,hu2022microscopic}.

On the other hand, one may argue that one main advantage of two-dimensional spectroscopy is the ability to access incoherent processes, such as dephasing, decay channels and inhomogeneous braodening.
In this sense, the purely Hamiltonian nature of our model is a limitation and inclusion of non-Hermitian elements is an interesting research direction.
In particular, incoherent processes  in real materials may cause pair breaking and  make the size of the quasiparticle gap of a superconductor
detectable in the cross-peak dynamics. Another valuable question is whether strong interactions may lead to complex two-dimensional spectra~\cite{valmilspild2022strongly}.

~

We are grateful to Atac Imamoglu and Jacek Kasprzak
for useful discussions.

\bibliography{bibliography}

\end{document}